\documentclass{article}
\usepackage{spconf,amsmath,graphicx,amssymb}
\usepackage{amsfonts} 
\usepackage{mathtools}
\usepackage{float}
\usepackage{mathptmx, mathrsfs}
\usepackage{afterpage}
\usepackage{caption}
\usepackage{subfigure}
\usepackage{color,xcolor}
\usepackage{multirow}
\usepackage{hyperref}
\usepackage{threeparttable}
\hypersetup{
    colorlinks=true,
    linkcolor=black,
    citecolor=black,
    urlcolor=black,
    linkbordercolor=white,
}

\def\@fnsymbol#1{\ensuremath{\ifcase#1\or *\or \dagger\or \ddagger\or
   \mathsection\or \mathparagraph\or \|\or **\or \dagger\dagger
   \or \ddagger\ddagger \else\@ctrerr\fi}}

\newcommand{\ssymbol}[1]{^{\@fnsymbol{#1}}}

\title{Implicit Regression in Subspace for High-Sensitivity CEST Imaging}
\name{Chu Chen$^{1,2}$, Yang Liu$^{2,3}$, Se Weon Park$^{2,3}$, Jizhou Li$^5$, Kannie W. Y. Chan$^{2,3,4}$, Raymond H. F. Chan$^{1,2}$\sthanks{Corresponding author (raymond.chan@cityu.edu.hk).}}
\address{
$^1$ Department of Mathematics, City University of Hong Kong, Hong Kong, China \\
$^2$ Hong Kong Centre for Cerebro-Cardiovascular Health Engineering, Hong Kong, China\\
$^3$ Department of Biomedical Engineering, City University of Hong Kong, Hong Kong, China\\
$^4$ Russell H. Morgan Department of Radiology and Radiological Science,\\ The Johns Hopkins University School of Medicine, Baltimore, MD, USA\\
$^5$ School of Data Science, City University of Hong Kong, Hong Kong, China}

\begin{document}
\ninept
\maketitle
\begin{abstract}
Chemical Exchange Saturation Transfer (CEST) MRI demonstrates its capability in significantly enhancing the detection of proteins and metabolites with low concentrations through exchangeable protons. The clinical application of CEST, however, is constrained by its low contrast and low signal-to-noise ratio (SNR) in the acquired data. Denoising, as one of the post-processing stages for CEST data, can effectively improve the accuracy of CEST quantification. In this work, by modeling spatial variant z-spectrums into low-dimensional subspace, we introduce \textit{Implicit Regression in Subspace (IRIS)}, which is an unsupervised denoising algorithm utilizing the excellent property of implicit neural representation for continuous mapping. Experiments conducted on both synthetic and in-vivo data demonstrate that our proposed method surpasses other CEST denoising methods regarding both qualitative and quantitative performance.
\end{abstract}
\begin{keywords}
CEST MRI, Denoising, Neural Network
\end{keywords}
\section{Introduction}
\label{sec:intro}
CEST (Chemical Exchange Saturation Transfer) is an imaging technique that detects diluted molecules by transferring their saturation to the abundant water pool. It has been reported that various diluted solutes, including peptides, proteins, and creatine~\cite{zhou2011differentiation,haris2014technique,chen2020vivo}, can be detected using CEST. Clinical interest in spectrally selective protein CEST effects has grown due to their potential in measuring and predicting brain tumor therapy response. CEST analysis involves studying protein-related CEST effects, such as amide proton transfer (APT) and relayed nuclear Overhauser effects (rNOE), by exploring small peaks in the z-spectra. However, the reliable detection of CEST effects is impeded by the inherently low signal-to-noise ratio (SNR) in CEST imaging~\cite{jones2013nuclear,zaiss2018chemical}. The diminished CEST contrast observed in tissues, as compared to solutions, is mainly due to factors such as direct water saturation (DS), magnetization transfer (MT) contrast, and the presence of various exchangeable protons from metabolites~\cite{zaiss2015combined}. These factors collectively contribute to the decreased sensitivity of CEST MRI quantification, thereby complicating the acquisition of accurate and reproducible measurements.

To fully unleash the potential of CEST imaging, denoising has gained attention as an effective post-processing strategy, where no additional data collection or equipment is required. A method based on multilinear singular value decomposition (MLSVD)~\cite{chen2020high} decomposed CEST data, leveraging spatiotemporal correlation and the low-rank property of CEST data. However, SVD-based methods with only truncation tend to preserve noise-contributing components, especially in severely corrupted data. Another approach, NLmCED~\cite{romdhane2021evaluation}, utilized the local self-similarity of CEST images to enhance SNR but failed to retain details within regions of interest and resulted in over-smoothing.

Apart from conventional algorithms, deep learning-based models have been applied to CEST denoising problems recently. For example, Chen \textit{et al.}~\cite{chen2023learned} proposed the Denoising CEST Network (DECENT) composed of two U-Nets as parallel branches, which aimed to extract spatiotemporal correlation and enhance spectral correlation respectively. However, DECENT requires many training samples and extensive optimization time for its dual U-Net architecture. Additionally, using a zero-padding strategy in DECENT to address inconsistent input sizes leads to considerable variation in output quality.

The CEST images acquired at various chemical shifts reveal a strong correlation associated exclusively with contrast alterations, which define the object's contour in the observed field. The discernible differences between these images stem from the distinct longitudinal signals that arise from the CEST effect across the chemical shifts. Based on this observation, it is possible to decompose noisy CEST data into a foundational low-dimensional structure along with its corresponding coefficients. By applying a non-linear threshold to these coefficients, we can enable the reconstruction of the images without the confounding presence of noise. Building on this concept, we developed the Implicit Regression in Subspace (IRIS) algorithm that incorporates singular value decomposition (SVD) and implicit neural representation~\cite{sitzmann2020implicit} to learn and map the continuous spatial coefficients for high-sensitivity CEST imaging. Validation of the IRIS algorithm on both phantom and in-vivo datasets has demonstrated its superior performance over existing CEST denoising techniques, significantly enhancing the SNR of the resulting z-spectra.

The main contributions are summarized as follows:
\begin{enumerate}
    \item We propose an unsupervised denoising algorithm (IRIS) for improving the SNR and sensitivity of CEST MR imaging.
    \item By reducing data dimensionality, IRIS leverages an extremely lightweight neural network while achieving high performance in representing continuous signals.
    \item Qualitative and quantitative evaluations on both synthetic and in-vivo datasets demonstrate the outperformance of IRIS over other methods in terms of noise reduction while maintaining the integrity of the underlying CEST signal.
\end{enumerate}

\begin{figure*}[!t]
\centering
  \includegraphics[height=0.2\textheight]{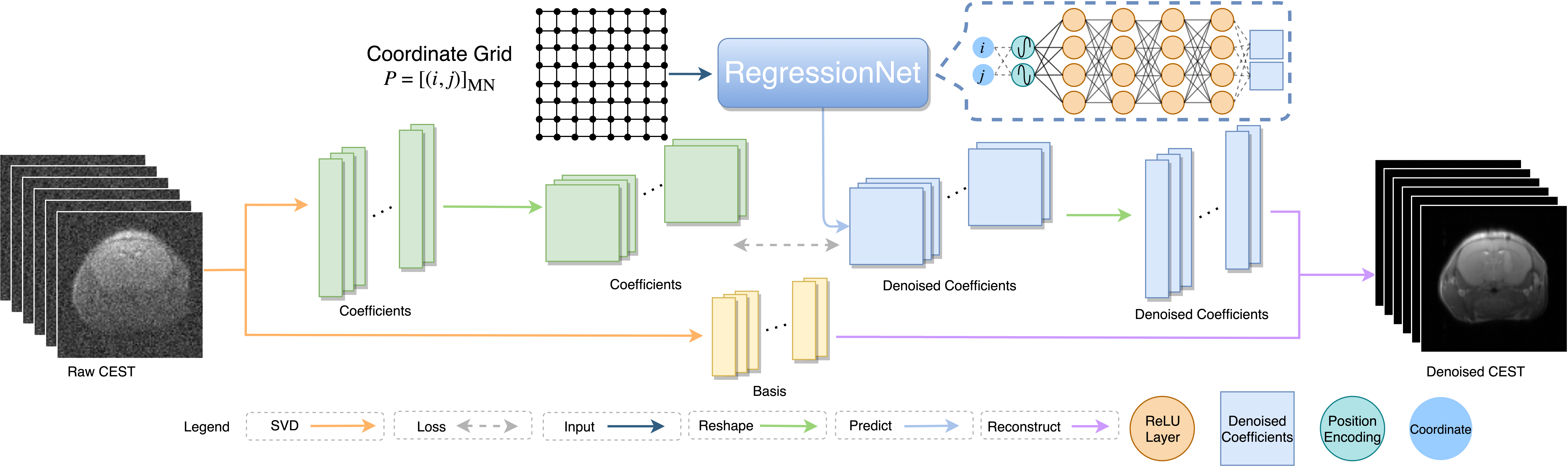}
  \vspace{-8pt}
  \caption{\textbf{IRIS Framework.} Input CEST data is decomposed into spatial coefficients and temporal basis by SVD. RegressionNet predicts the coefficients for each position based on the coordinates input and is optimized through a loss function calculated by comparing predicted coefficients with those from SVD. Denoised CEST data is then reconstructed from predicted coefficients and temporal basis.}
  \label{IRISStruct}
  \vspace{-10pt}
\end{figure*}

\section{Method}
\vspace{-5pt}
\label{sec:method}

During the acquisition process, the CEST data is stored in k-space, which represents the spatial frequency domain and serves as the basis for image reconstruction. CEST noise is initially introduced in k-space, but due to the linear transformation (e.g. Foriour transform) from k-space to the image domain, it is equivalent to introducing additional noise in the image domain. With the additive noise assumption for CEST data in $M\times N$ spatial dimension and number of $C$ offsets, the data formation model can be expressed as
\vspace{-3pt}
\begin{equation}
\label{eq1}
    \mathbf{y} = \mathbf{x} + \mathbf{n},
\vspace{-5pt}
\end{equation}
where $\mathbf{x}, \mathbf{y} \in \mathbb{R}^{MN\times C}$ are vectorized matrices of a noise-free and noisy pair in the real-valued image domain, respectively, while $\mathbf{n} \in \mathbb{R}^{MN\times C}$ is the noise matrix.

To obtain the underlying $\mathbf{x}$ from measured data $\mathbf{y}$, we introduce a novel algorithm: Implicit Regression in Subspace (IRIS). The overall CEST denoising procedure of IRIS is displayed in Fig.~\ref{IRISStruct}. Two primary components of IRIS are detailed in the subsequent sections.
\vspace{-15pt}
\subsection{Subspace Formulation}
The property of spatiotemporal correlation in CEST imaging indicates that high-dimensional spectroscopic signals can be reconstructed from a low-dimensional subspace~\cite{chen2020high,lam2014subspace}. Noise, however, possessed no spatial or ``temporal" correlation compared to actual CEST effects. To this end, we can decompose the noise-free data $\mathbf{x}$ into $K$-dimensional pairs of components as $\mathbf{x}=\mathbf{u}\mathbf{v}$, where $\mathbf{v} \in \mathbb{R}^{K\times C}$ represents basis with respect to chemical shifts and $\mathbf{u} \in \mathbb{R}^{MN\times K}$ stores the corresponding spatial dependent coefficients. This decomposition can be accomplished through singular value decomposition (SVD) of the measured data $\mathbf{y}$.

By adopting the concept of transformation-based denoising methods~\cite{blu2007sure,li2023sum}, we can remove the components related to noise from the coefficients $\mathbf{u}$ such that noise-free data $\mathbf{x}$ can be reconstructed with basis $\mathbf{v}$. Therefore, denoising problem expressed in Eq.~(\ref{eq1}) can now be reformulated into an optimization problem
\vspace{-3pt}
\begin{equation} \label{eq:de-form}
    \operatorname*{argmin}_{\mathbf{u}} \frac{1}{2} || \mathbf{uv}-\mathbf{y} ||^2 + \lambda R(\mathbf{u}),
\vspace{-8pt}
\end{equation}
where $R(\mathbf{u})$ is regularization term varied with prior of $\mathbf{u}$. While $\mathbf{v}$ is orthogonal matrix, Eq.~(\ref{eq:de-form}) is equivalent to
\begin{equation}\label{eq:de-form-equi}
\operatorname*{argmin}_{\mathbf{u}} \frac{1}{2} || \mathbf{u}-\mathbf{y}\mathbf{v}^{T} ||^2 + \lambda R(\mathbf{u}),
\vspace{-8pt}
\end{equation}

$R(\mathbf{u})$ can be approximated by the summation of elementary functions $R_{k}(\mathbf{u}^{k})$. Thus, the solution to Eq.~(\ref{eq:de-form-equi}) can be re-written in a vectorized form as
\begin{equation}
\hat{\mathbf{u}} = [\hat{\mathbf{u}}^{1};\dots \hat{\mathbf{u}}^{k};\dots \hat{\mathbf{u}}^{K}],
\hat{\mathbf{u}}^{k} = \textit{D}_{\lambda}^{k}(\mathbf{y}\mathbf{v}_{k}^{T}),
\vspace{-8pt}
\end{equation}
where $\textit{D}_{\lambda}^{k}$ is thresholding operator which can be defined as
\vspace{-3pt}
\begin{equation}
\label{eq:sub-prob}
\textit{D}_{\lambda}^{k}(\mathbf{w}) = \operatorname*{argmin}_{\mathbf{u}} \frac{1}{2} || \mathbf{u}-\mathbf{w} ||^2 + \lambda R_{k}(\mathbf{u}).
\vspace{-8pt}
\end{equation}

As Eq.~(\ref{eq:sub-prob}) can be regarded as a typical image denoising problem, the original CEST denoising problem can now be addressed by solving a total of $K$ image denoising sub-problems.
\vspace{-8pt}
    \begin{figure}
    \centering
        \includegraphics[height=0.2\textwidth]{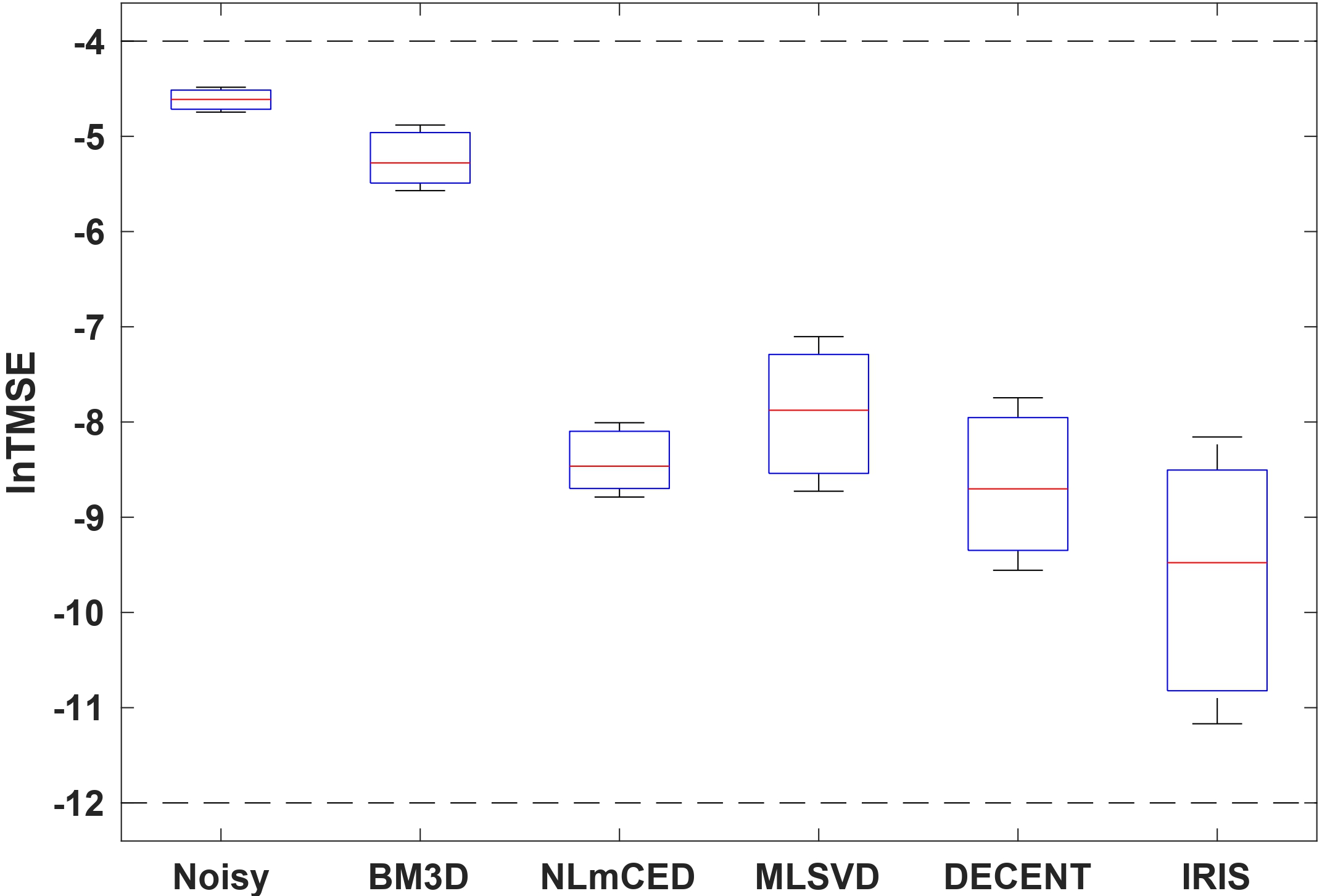}
        \vspace{-5pt}
        \caption{lnTMSE shown in boxplots indicates the median and inter-quartile range.}
        \label{phan_stat}
        \vspace{-10pt}
    \end{figure}
    \begin{figure*}[!t]
    \subfigure[Noisy]{
        \label{noisy_phan}
        \includegraphics[width=0.136\linewidth]{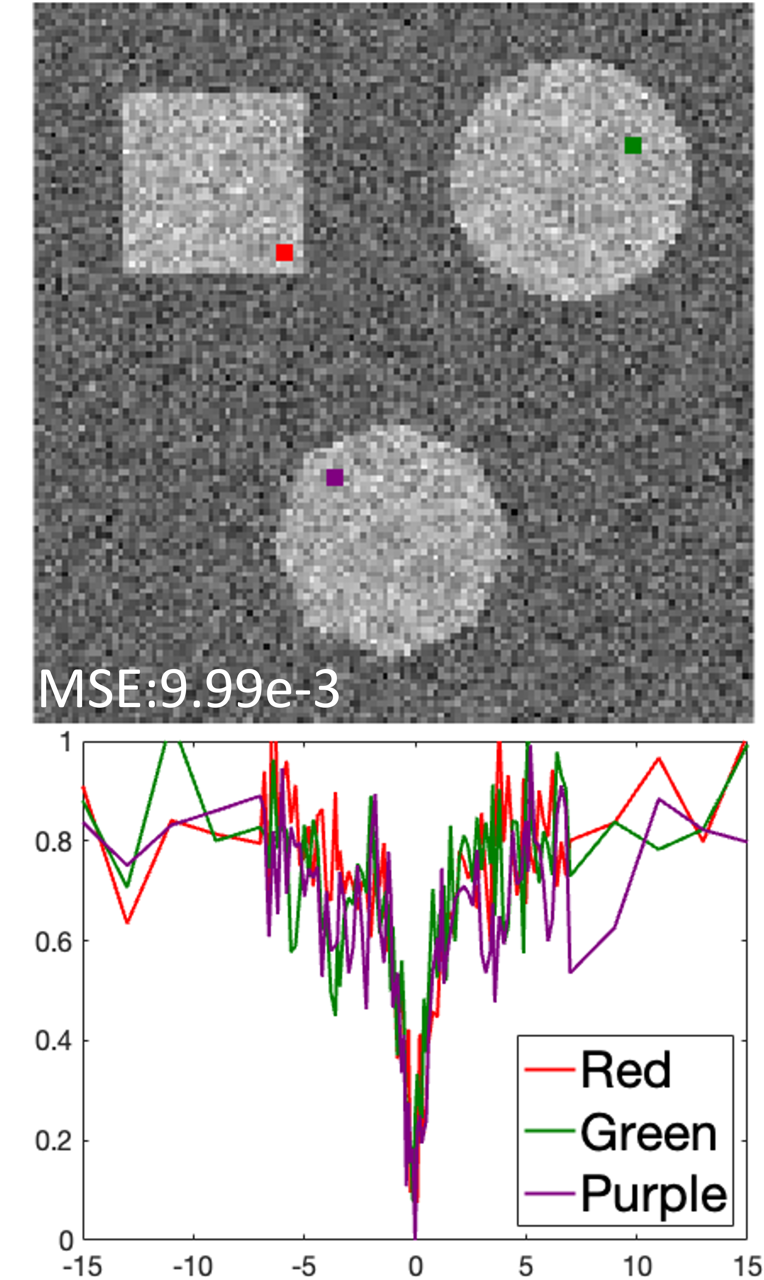}}
    \subfigure[BM3D~\cite{dabov2007image}]{
        \label{BM3D_phan}
        \includegraphics[width=0.136\linewidth]{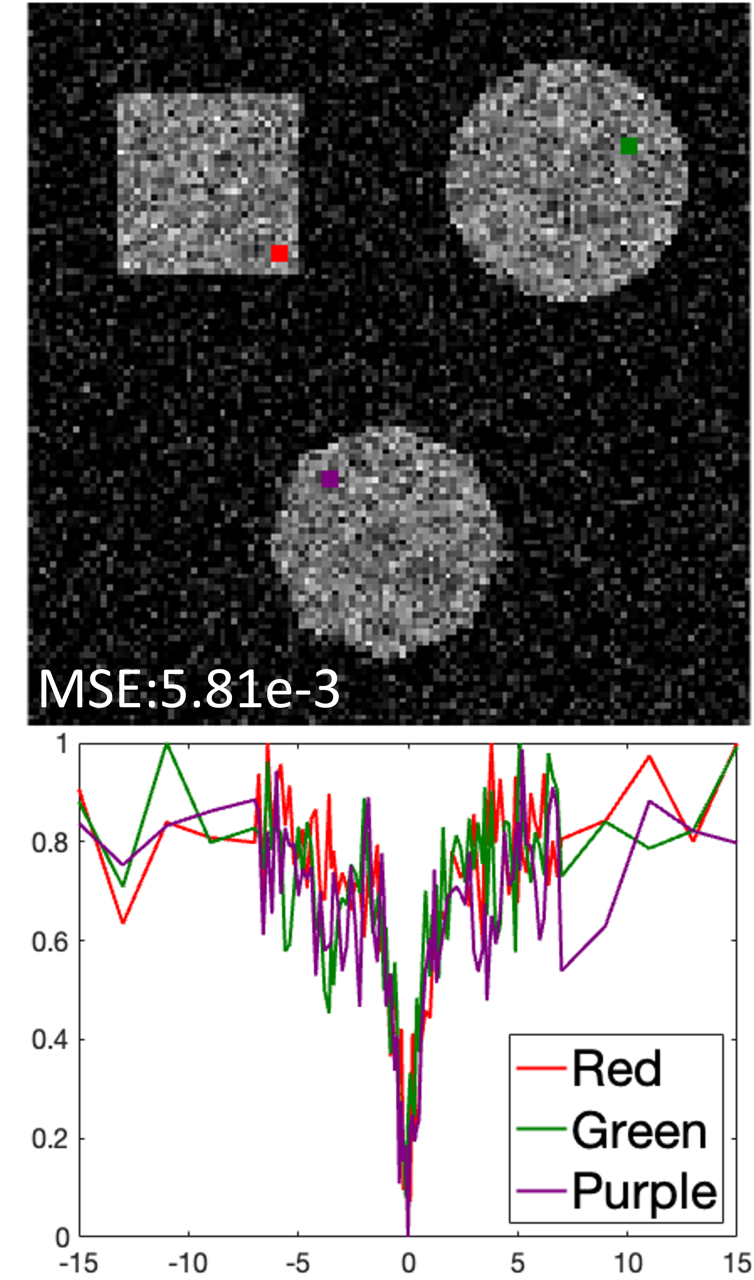}}
    \subfigure[NLmCED~\cite{romdhane2021evaluation}]{
        \label{NLmCED_phan}
        \includegraphics[width=0.136\linewidth]{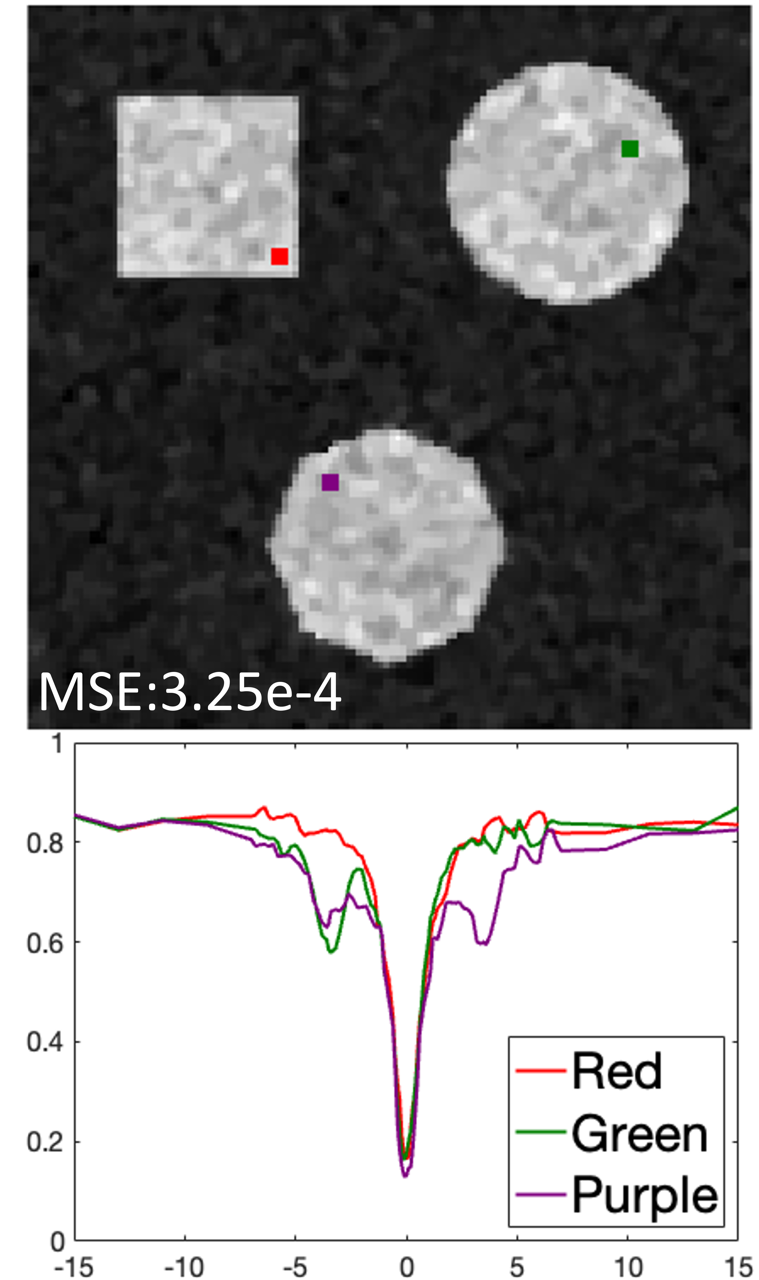}}
    \subfigure[MLSVD~\cite{chen2020high}]{
        \label{mlsvd_phan}
        \includegraphics[width=0.136\linewidth]{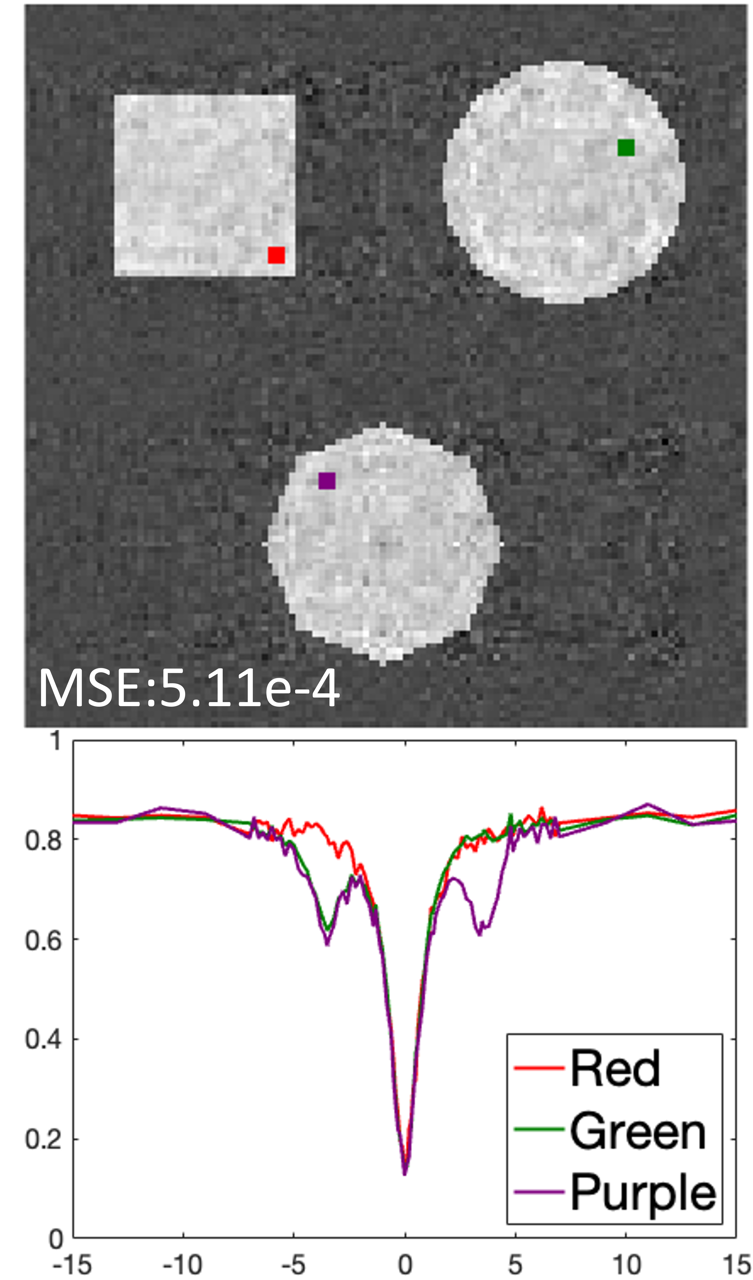}}
    \subfigure[DECENT~\cite{chen2023learned}]{
        \label{DECENT_phan}
        \includegraphics[width=0.136\linewidth]{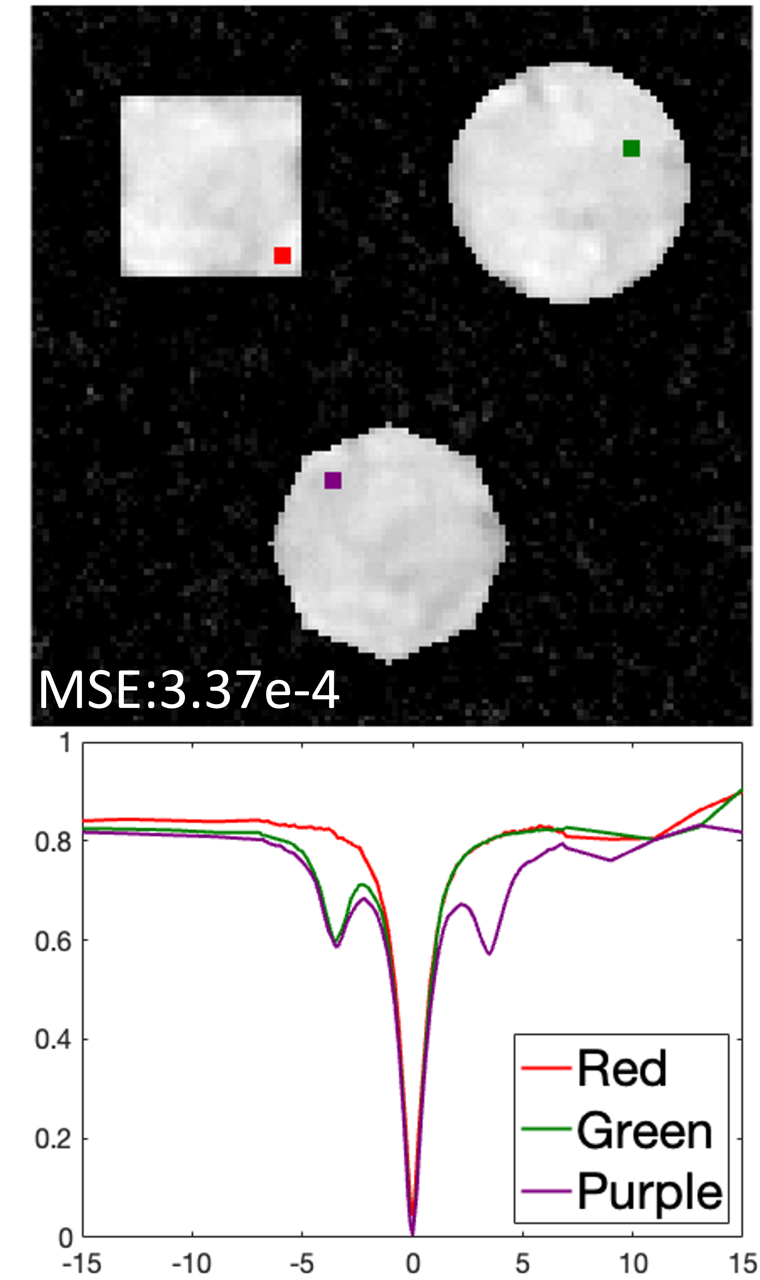}}
    \subfigure[IRIS~(\textbf{Ours})]{
        \label{IRIS_phan}
        \includegraphics[width=0.135\linewidth]{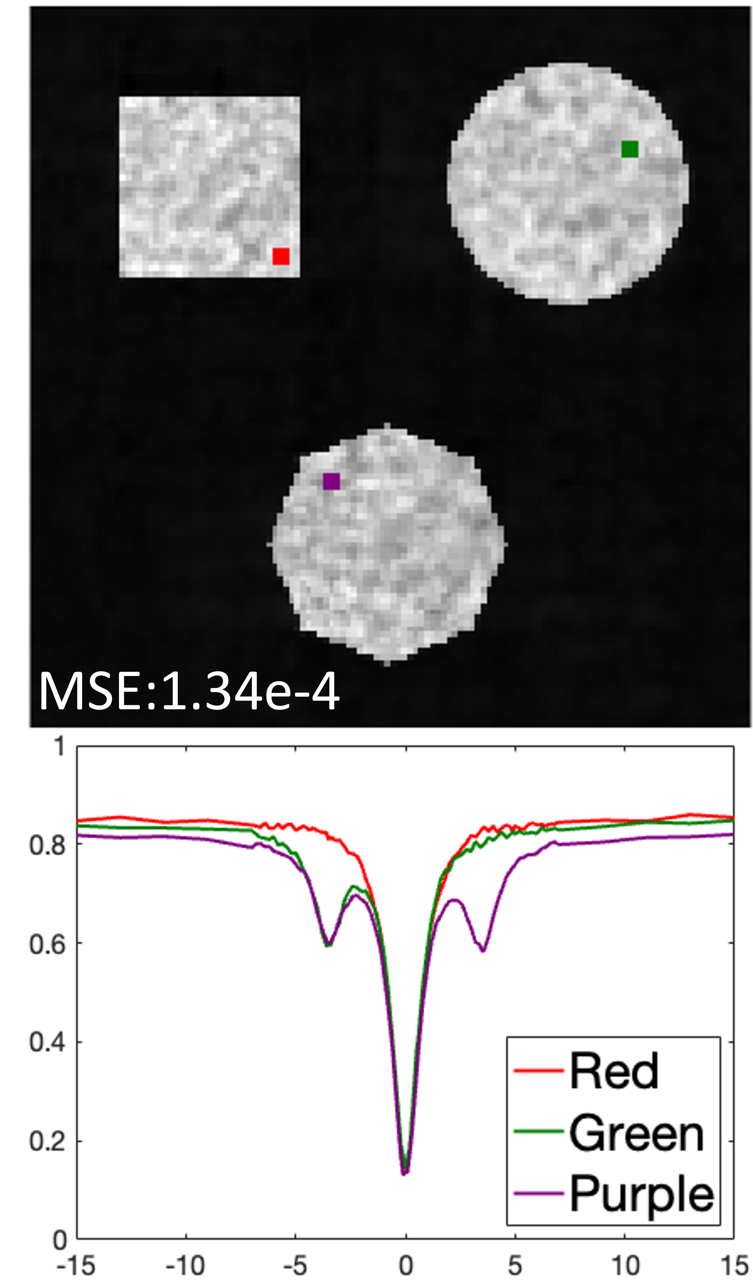}}
    \subfigure[Ground Truth]{
        \label{GT_phan}
        \includegraphics[width=0.131\linewidth]{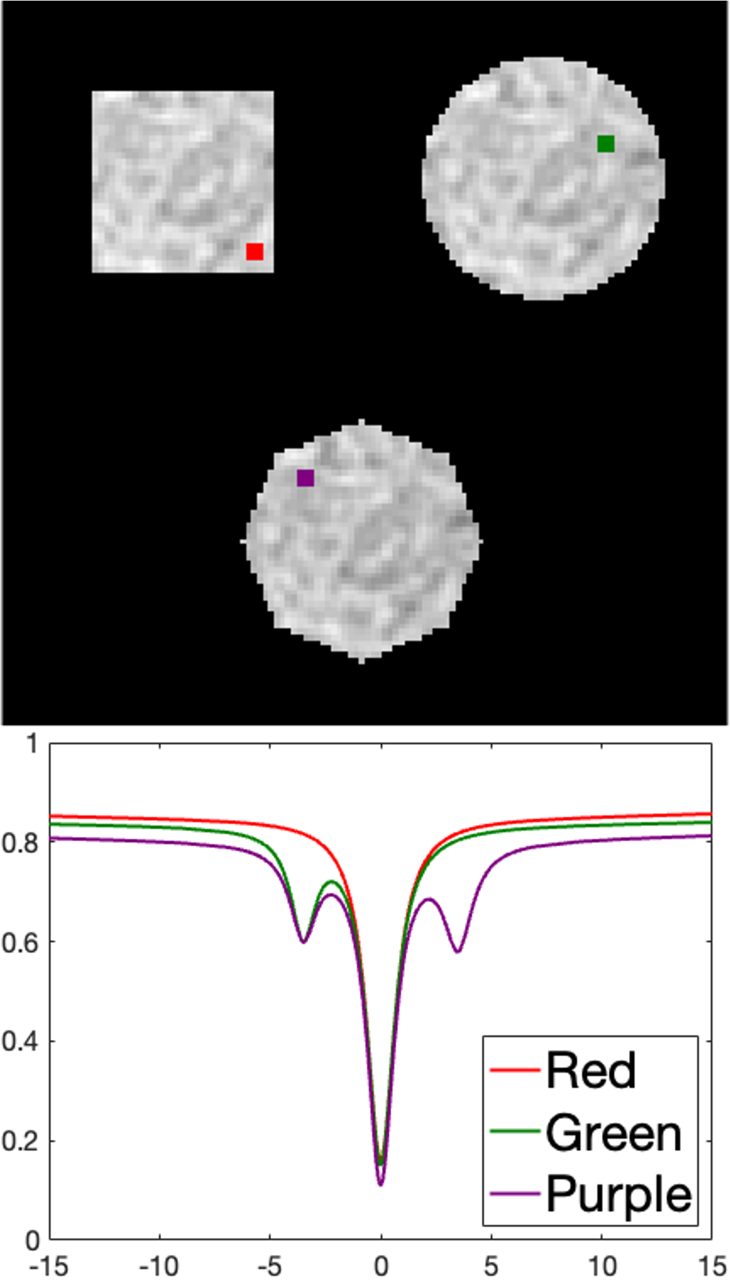}}
    \vspace{-10pt}
    \caption{Denoised phantom data and signals located in different phantoms indicated by color dots.}
    \vspace{-10pt}
    \label{phan}
    \end{figure*}

    \begin{figure*}[!ht]
    \centering
    \subfigure[Noisy]{
        \label{noisy_nod7}
        \includegraphics[width=0.14\linewidth]{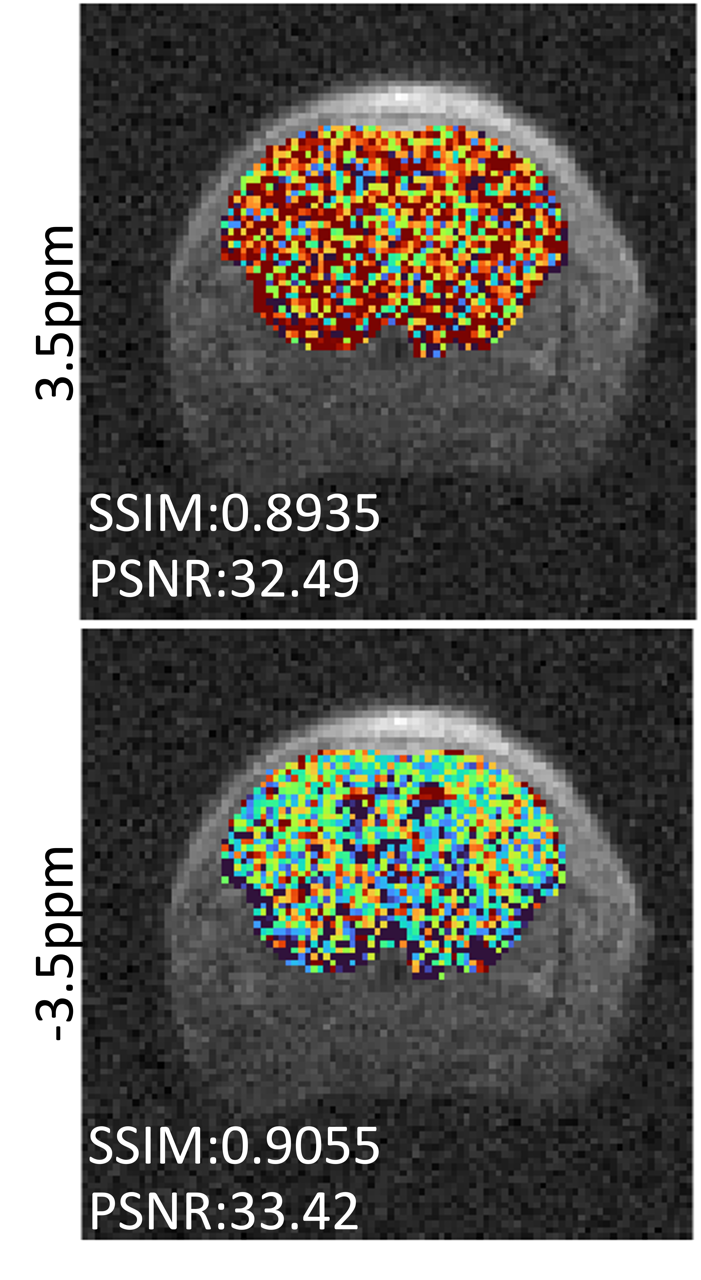}}
    \subfigure[NLmCED~\cite{romdhane2021evaluation}]{
        \label{NLmCED_nod7}
        \includegraphics[width=0.13\linewidth]{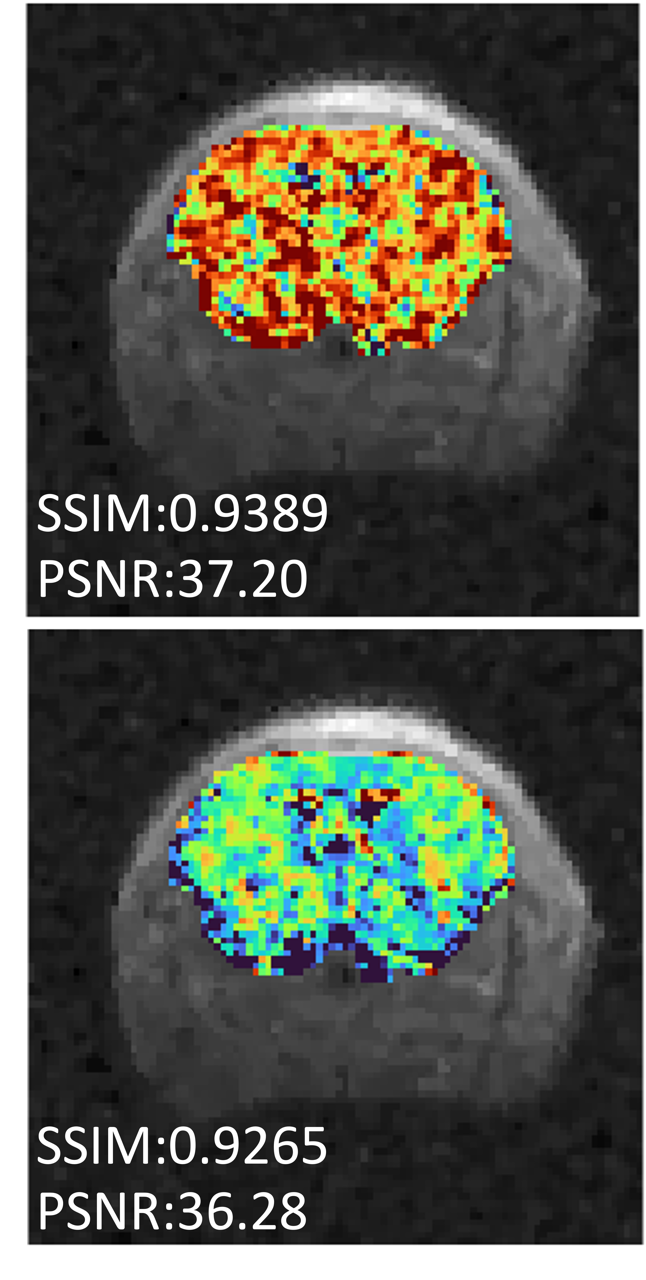}}
    \subfigure[MLSVD~\cite{chen2020high}]{
        \label{mlsvd_nod7}
        \includegraphics[width=0.13\linewidth]{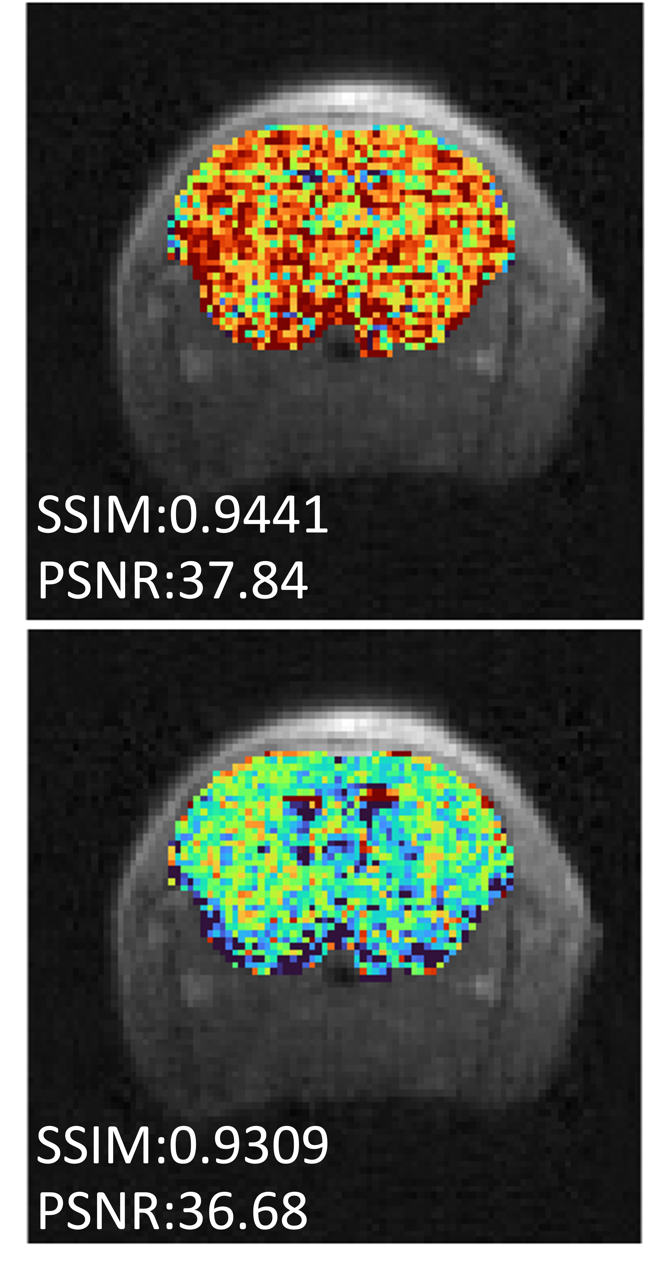}}
    \subfigure[DECENT~\cite{chen2023learned}]{
        \label{DECENT_nod7}
        \includegraphics[width=0.13\linewidth]{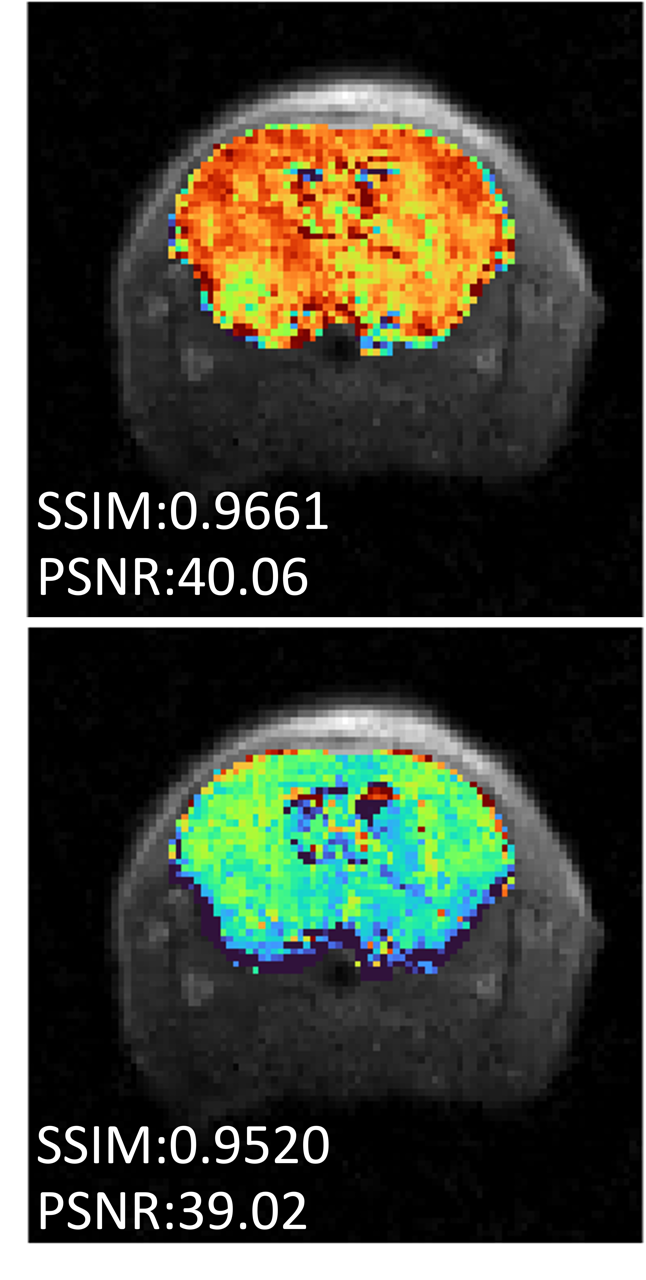}}
    \subfigure[IRIS~(\textbf{Ours})]{
        \label{IRIS_nod7}
        \includegraphics[width=0.131\linewidth]{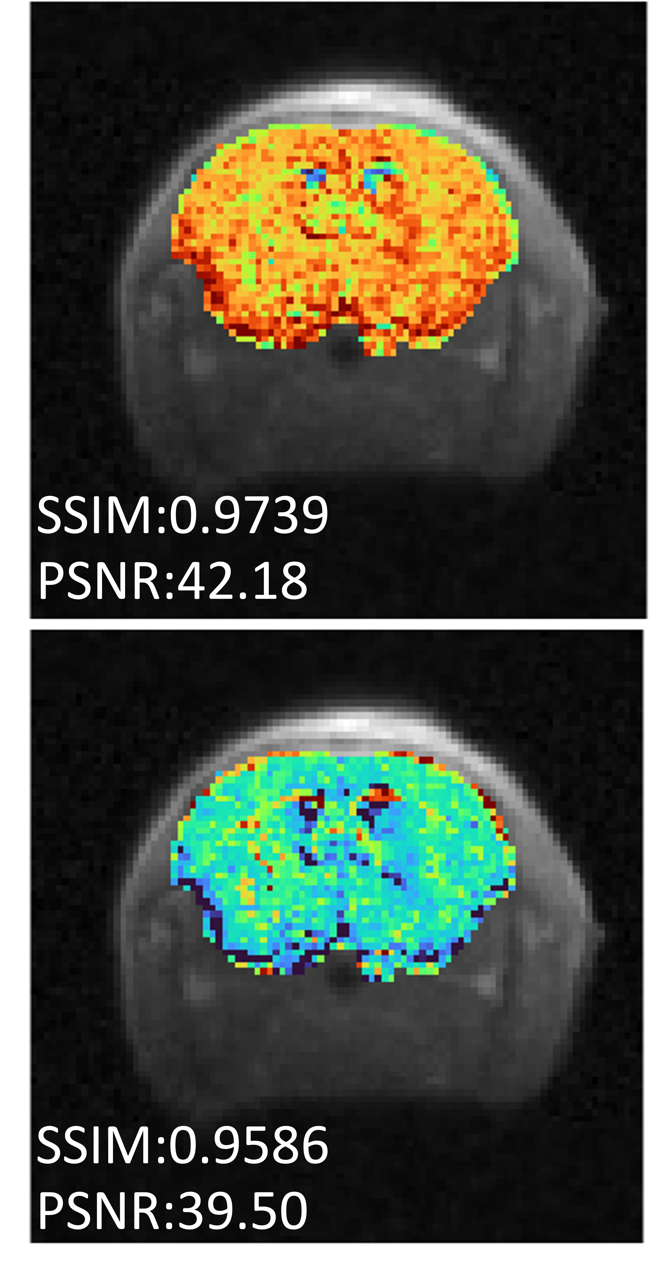}}
    \subfigure[Reference]{
        \label{GT_nod7}
        \includegraphics[width=0.151\linewidth]{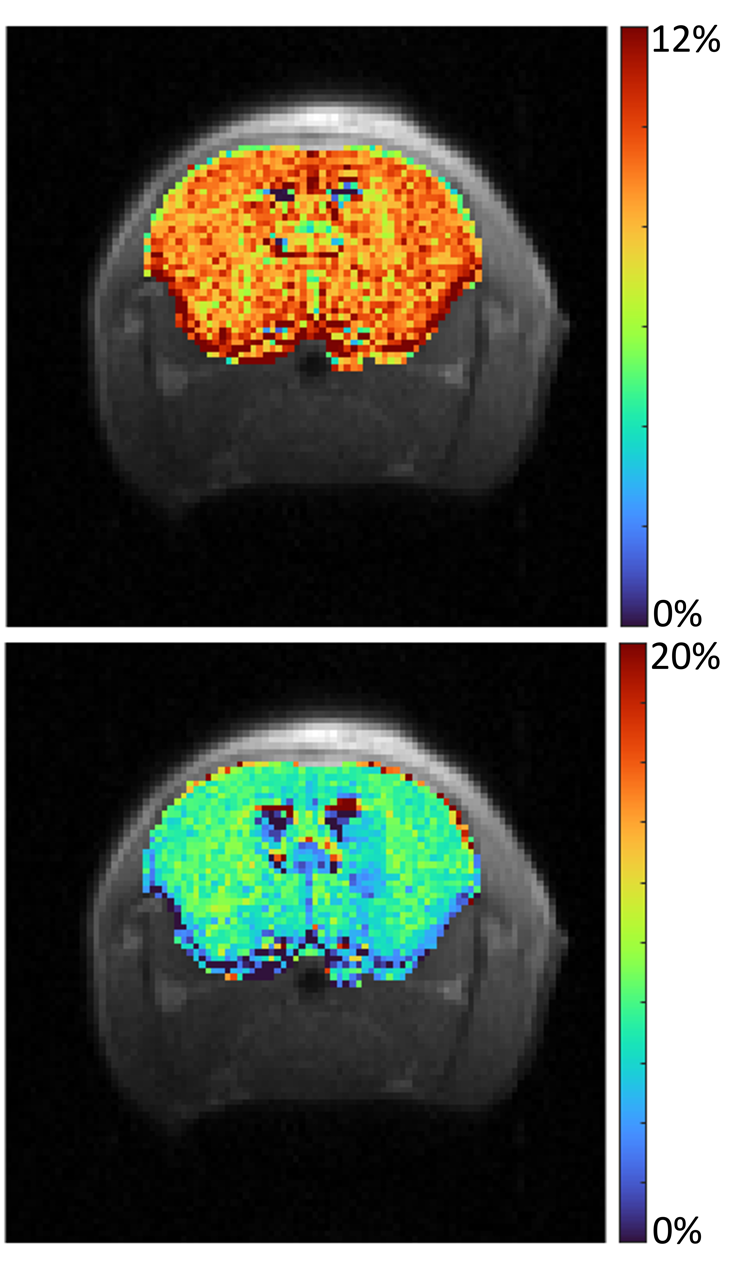}}
    \subfigure[T2w]{
        \label{T2w_nod7}
        \includegraphics[width=0.12\linewidth]{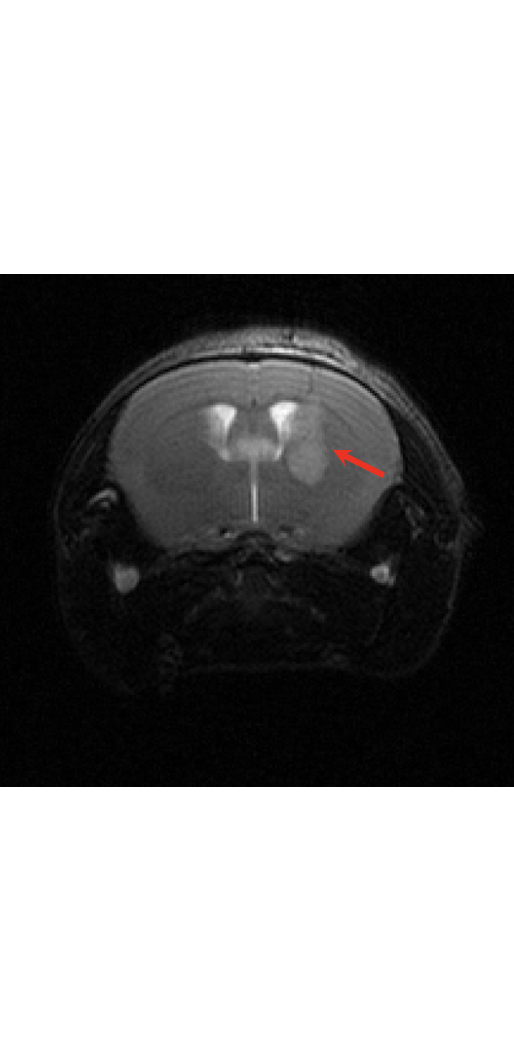}}
    \vspace{-8pt}
    \caption{APT (upper) and rNOE (lower) mapping from clean subject with additive noise ($\sigma=0.05$). The red arrow in T2w image indicates the tumor region.}
    \label{nod7}
    \vspace{-10pt}
    \end{figure*}
\vspace{-2pt}
\subsection{Implicit Regression}
\vspace{-2pt}
Deep implicit neural network has shown strong capability of representing complicated continuous functions~\cite{mildenhall2021nerf,sitzmann2020implicit,Saragadam_2023_CVPR}, including image representation. Knowing the fact that outliers in the given data are overlooked when it is fitted by a smooth continuous function (\textit{i.e.} regression), a denoising result can be obtained by taking the regression curve. Therefore, once $K$ has been determined, IRIS finds a neural network called RegressionNet $\textit{N}_{\theta}$ to represent continuous mapping $f: \mathbb{R}^{MN\times 2} \rightarrow \mathbb{R}^{MN\times K}$ that takes coordinate grid $P=[(i,j)]_{MN}$ as input and aims to solve $K$ denoising problems in Eq.~(\ref{eq:sub-prob}) simultaneously as $K$-dimension can be regarded as parallel channels
\begin{equation}
\label{eq:network-opt}
 \hat{\theta} = \operatorname*{argmin}_{\theta} \frac{1}{2} || \textit{N}_{\theta}(P)-\mathbf{y}\mathbf{v}^{T} ||^2,
\vspace{-5pt}
\end{equation}
such that only one network is required, resulting in algorithm efficiency and less burden on defining hyper-parameters.

In the case of CEST denoising, backpropagation of the RegressionNet is realized by minimizing the following loss function,
\begin{equation}
\label{eq:loss}
\textit{L} = || \textit{N}_{\theta}(P)-\mathbf{y}\mathbf{v}^{T} ||_{2},
\end{equation}
$\ell_2$-norm is applied here, preventing over-fitting after regression. Once optimization is finished, denoised CEST data can be reconstructed by
\begin{equation}
    \hat{\mathbf{x}} = \hat{\mathbf{u}}\mathbf{v} = \textit{N}_{\hat{\theta}}(P)\mathbf{v}.
\vspace{-8pt}
\end{equation}

\section{Experiments}
\vspace{-10pt}
\label{sec:experiments}
\subsection{Data Setup}
\vspace{-6pt}
\textit{Synthetic Phantom.} We simulated three sets of z-spectra by varying the number of basis functions in the Lorentzian model~\cite{zaiss2013chemical,goerke2019relaxation}. The square phantom represented the direct water saturation with the MT effect, the circular phantom included the extra rNOE effect, and the octagonal phantom added the APT on top of the previous two effects. In addition, we applied Gaussian filtering to the randomly generated parameters in the Lorentzian model to mimic spatial variance and smoothness (find \href{https://htmlpreview.github.io/?https://github.com/chuchen1206/IRIS_CEST-Denoising/blob/9e28e72a48ef32459931e468a6526df3808e03e0/IRIS_Appendix.html}{Appendix} for simulation details).

\textit{Animal Experiment.} For in-vivo evaluation, we have prepared two sets of mice data with tumors: one with minimal noise level (\textit{Clean Subject}) and the other with more noticeable noise (\textit{Noisy Subject}). Both were injected with U-87 MG cell ($0.5 M/ 3ul$) at 2.0 mm right-lateral, 0.2 mm anterior, and 3.8 mm below the bregma. The CEST MRI sequence was a continuous-wave (CW) saturation module followed by the rapid acquisition with refocused echoes (RARE) as a readout module. A power (B1) of $0.8 \mu T$ and a duration ($t_{sat}$) of 3000 ms were used for the saturation module. The saturation frequency varied from $–15$ to 15 ppm, with a 0.2 ppm increment and a 2 ppm increment between –7/7 and –15/15 ppm. Four M0 images with saturation frequency offset at 200 ppm were acquired and averaged for z-spectrum normalization. The readout parameters were as follows: repetition time (TR)=5000 ms, echo time (TE)=5.9 ms, matrix size=$96\times96$, slice thickness=1 mm, RARE factor=32.
\vspace{-15pt}

    \begin{figure*}[!t]
    \centering
    \subfigure[Noisy]{
        \label{noisy_nod5}
        \includegraphics[width=0.14\linewidth]{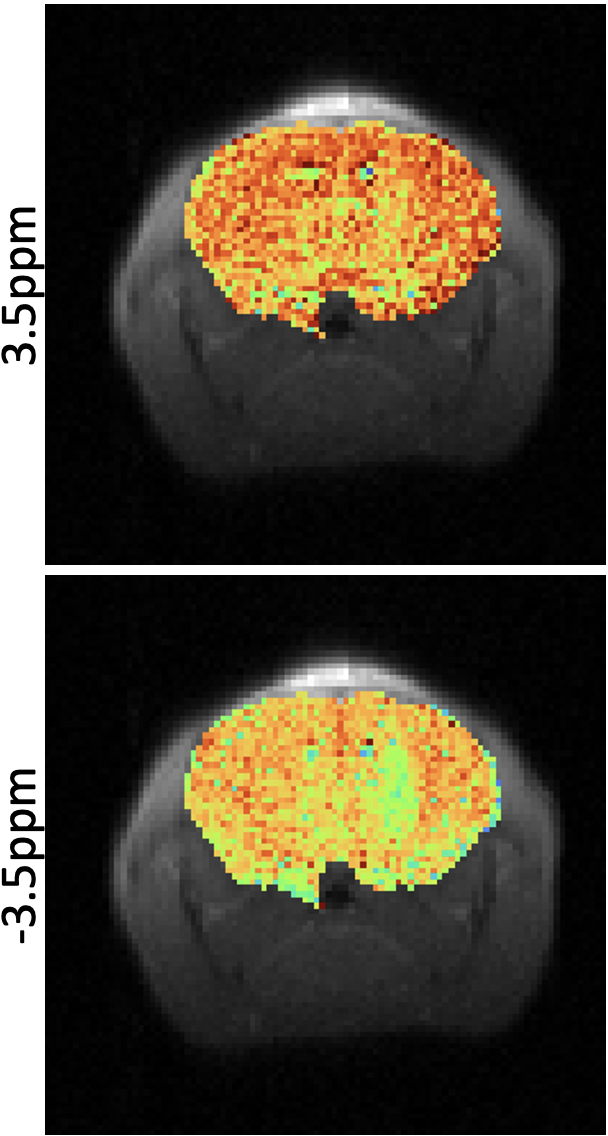}}
    \subfigure[NLmCED~\cite{romdhane2021evaluation}]{
        \label{NLmCED_nod5}
        \includegraphics[width=0.13\linewidth]{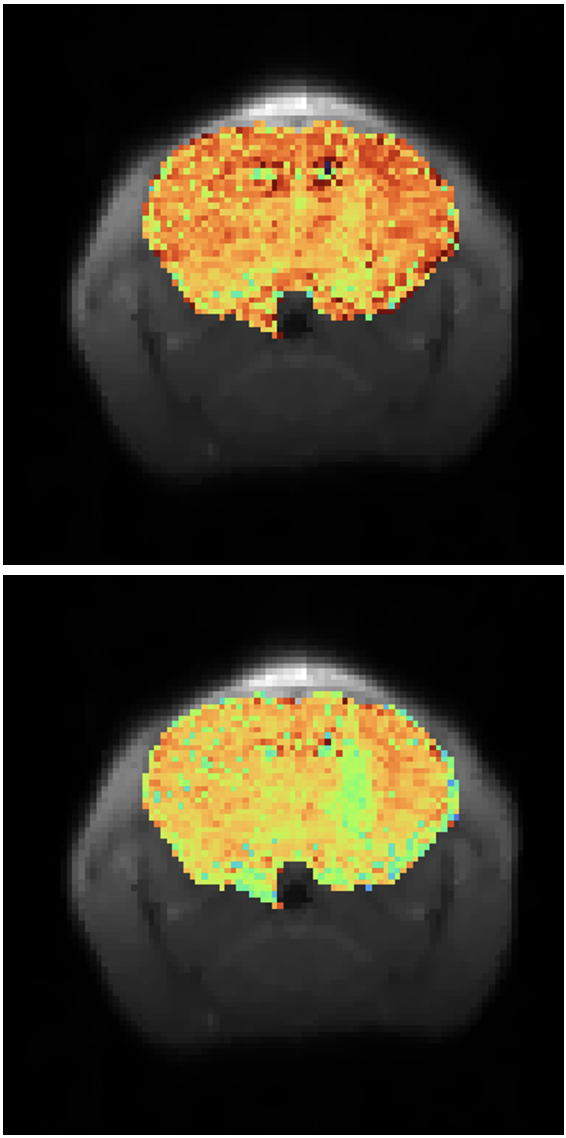}}
    \subfigure[MLSVD~\cite{chen2020high}]{
        \label{mlsvd_nod5}
        \includegraphics[width=0.13\linewidth]{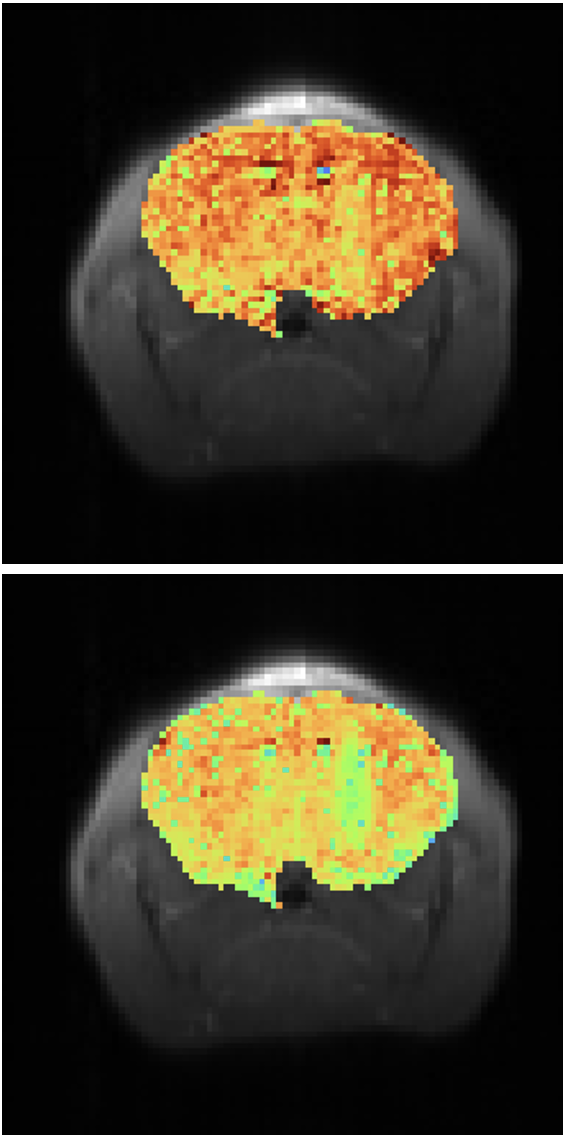}}
    \subfigure[DECENT~\cite{chen2023learned}]{
        \label{DECENT_nod5}
        \includegraphics[width=0.13\linewidth]{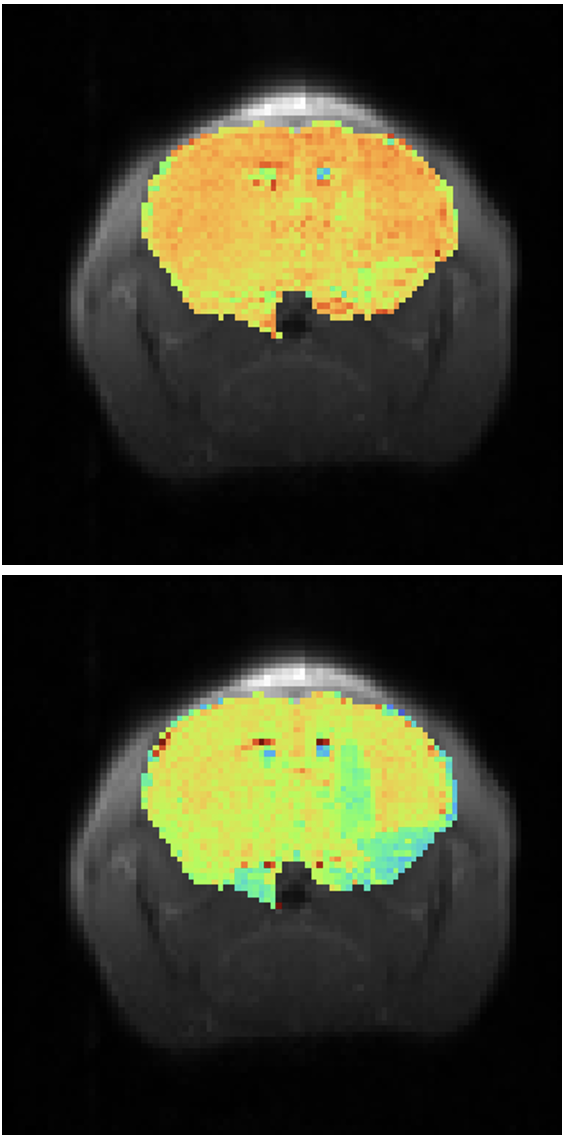}}
    \subfigure[IRIS~(\textbf{Ours})]{
        \label{IRIS_nod5}
        \includegraphics[width=0.15\linewidth]{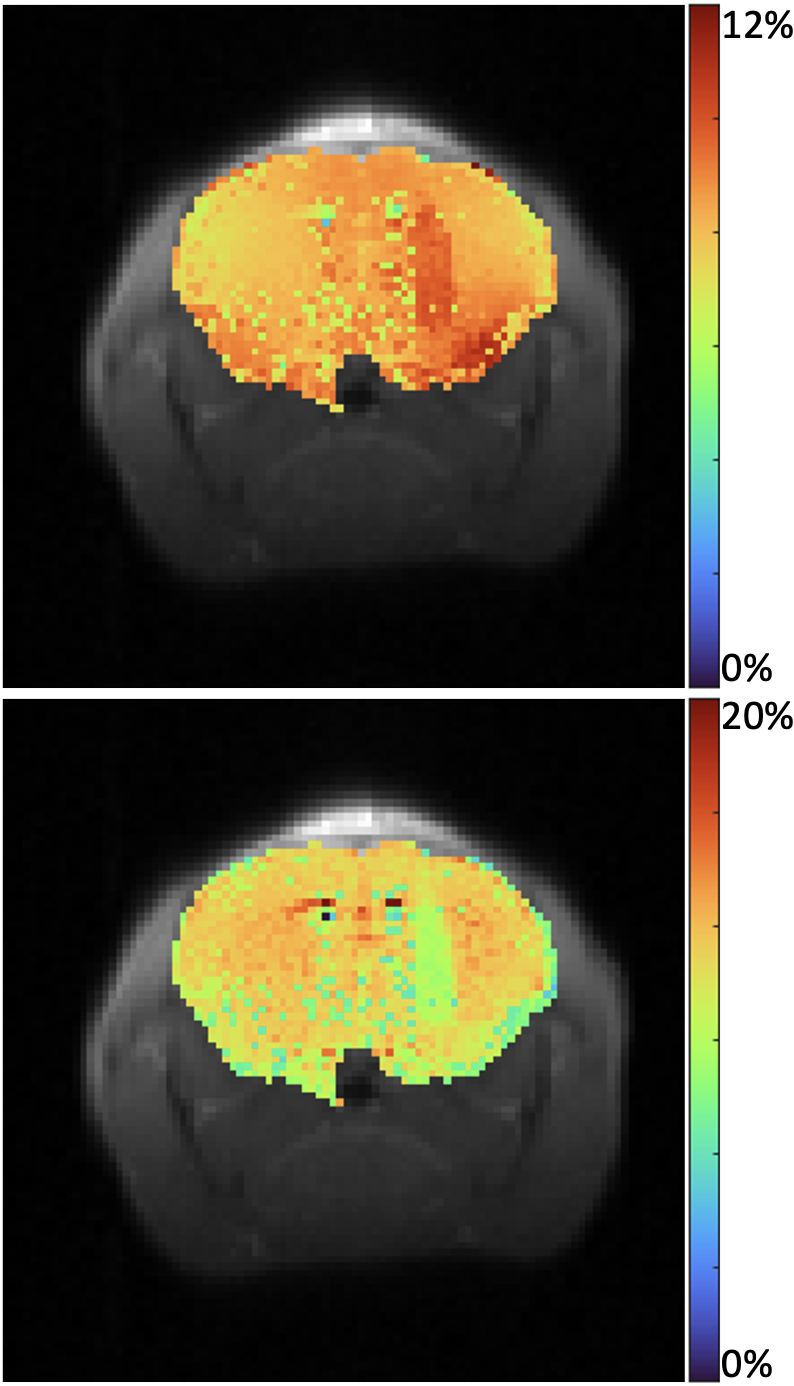}}
    \subfigure[T2w]{
        \label{T2w_nod5}
        \includegraphics[width=0.12\linewidth]{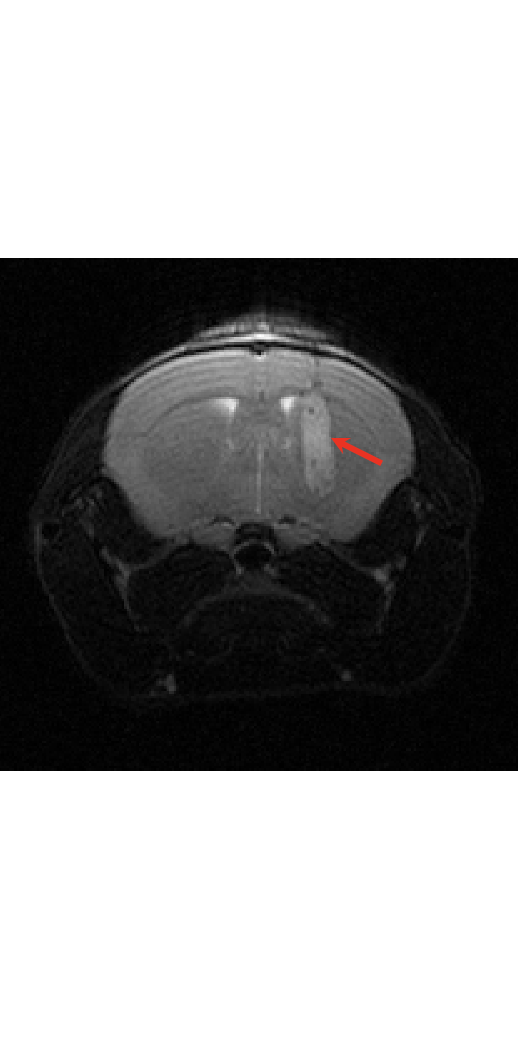}}
    \vspace{-8pt}
    \caption{APT (upper) and rNOE (lower) Mapping from noisy subject. The red arrow in T2w image indicates the tumor region.}
    \label{nod5}
    \vspace{-8pt}
    \end{figure*}

\subsection{Evaluations}
\vspace{-3pt}
We evaluate the proposed IRIS algorithm along with several state-of-the-art CEST denoising methods for comparison, including BM3D~\cite{dabov2007image}, NLmCED~\cite{romdhane2021evaluation}, MLSVD~\cite{chen2020high} and DECENT~\cite{chen2023learned}. Default parameter settings suggested by authors were adopted. BM3D is applied channel by channel.

Specifically for the architecture of RegressionNet, we use a position encoding layer~\cite{mildenhall2021nerf} for input and 4 linear layers with ReLU activation function and 512 neurons for feature extraction as highlighted in the dotted bounding box of Fig.~\ref{IRISStruct}. While other structures of INR are also applicable here, we use RegressionNet architecture for a faster iteration speed. $K=4$ and $1\times 10^{-3}$ of learning rate were used for 3000 iterations in practice. All experiments are conducted in Matlab (R2023a) or Python 3.8.17 on a PC equipped with Intel(R) Core(TM) i7-9700 CPU 3.00GHz and Nvidia GeForce RTX 3090 GPU with 24G of memory.
\vspace{-12pt}

\subsubsection{Evaluation Metrics}
\vspace{-5pt}
Mean square error (MSE), peak signal-to-noise ratio (PSNR), and structural similarity index (SSIM) were used for data fidelity and image-based quality assessment, and we additionally introduced natural logarithm of temporal MSE (lnTMSE) to evaluate voxel-wise error,
\begin{equation}
    \mathrm{lnTMSE}_{p} = \ln{\sum_{q=1}^{C} (\mathbf{x}_{pq} - \hat{\mathbf{x}}_{pq})^2}, p=1,\dots, MN.
\vspace{-10pt}
\end{equation}

\subsubsection{Synthetic Phantom}
\vspace{-5pt}
To create a noisy sample, the synthetic phantom was added with additional noise $\mathbf{n}\sim \mathcal{N}(0, 0.1^2)$. Fig.~\ref{phan_stat}, displaying the statistical distribution of lnMSE, illustrates all denoising approaches managed to improve data quality as lnMSEs were generally reduced. Among all, however, IRIS exhibited excellent performance with not only the lowest overall error but also outperforms other methods by having more than half of the signal errors smaller than their minimum error. CEST image at 0.4 ppm is shown in the upper row of Fig.~\ref{phan} demonstrating the denoising effect in spatial perspective, and the lower row plotted denoised signals at three locations of different classes as indicated in the image. Lacking finding a spatiotemporal correlation, BM3D failed to provide clean signals. While both NLmCED and MLSVD effectively preserve spatial smoothness, they still exhibit residual noise in the signal. DECENT often introduces significant errors in the denoised signal, particularly at endpoints and peaks. IRIS achieves a better result, as evidenced by its superior signal and spatial texture fidelity compared to the ground truth.
\vspace{-12pt}

\subsubsection{In-vivo Data}
\vspace{-8pt}
\textit{Clean Subject.} To mimic multi-level data quality, we introduced three levels of Gaussian noise, namely Mild ($\sigma = 0.05$), Medium ($\sigma = 0.1$), and Severe ($\sigma = 0.3$), to the clean CEST subject. Denoising performance from different methods is shown in Table~\ref{table:denoising results}. The SSIM and PSNR metrics demonstrate that IRIS performs exceptionally better than other baselines across multiple noise conditions. Meanwhile, we performed 4-pool Lorentzian fitting~\cite{goerke2019relaxation,glang2020deepcest} to denoising results for APT (3.5ppm) and rNOE (-3.5 ppm) maps (find \href{https://htmlpreview.github.io/?https://github.com/chuchen1206/IRIS_CEST-Denoising/blob/9e28e72a48ef32459931e468a6526df3808e03e0/IRIS_Appendix.html}{Appendix} for fitting details) as displayed in Fig.~\ref{nod7}. The presence of noise leads to significant errors in traditional fitting methods. Whereas the introduction of denoising techniques greatly improves the accuracy of the fitting results. The bottom-left corner of the image displays the comparison of SSIM and PSNR based on the generated CEST maps. IRIS has a strong capability of preserving the CEST effect from noisy data, as suggested by numerical and visual results. Furthermore,  as the tumor region possesses higher cell density, proliferation will lead to higher amide protons concentrations, whereas rNOE shows hypo-intensity~\cite{cai2015cest}. IRIS even suppressed the subtle noise in the clean subject, resulting in high intensity at the tumor region in the APT map.

    \begin{table}
    \begin{center}
    \caption{Performance evaluation by comparing denoising results from different methods to the ground truth (PSNR (dB)~/~SSIM). }
    \vspace{-5pt}
    \label{table:denoising results}
    \begin{threeparttable}
    \setlength{\tabcolsep}{2mm}{
    \begin{tabular}{c c c c}
    \hline   \textbf{Noise Level} & Mild & Medium & Severe \\
    \hline   Noisy & 26.02~/~0.7034 & 20.00~/~0.4023 & 10.46~/~0.0901 \\
            BM3D~\cite{dabov2007image} & 27.69~/~0.7562 & 21.69~/~0.4429 & 12.85~/~0.1034\\
            MLSVD~\cite{chen2020high} & 35.17~/~0.9498 & 30.18~/~0.8572 & 20.91~/~0.4317 \\
            NLmCED~\cite{romdhane2021evaluation}& 35.10~/~0.9592 & 32.64~/~0.9229 & 24.32~/~0.5954\\
            DECENT~\cite{chen2023learned} & 32.30~/~0.8343 & 25.12~/~0.7114 & 16.10~/~0.4258 \\
            IRIS(\textbf{Ours}) & \textbf{36.02}~/~\textbf{0.9635} & \textbf{33.58}~/~\textbf{0.9388} & \textbf{27.87}~/~\textbf{0.8385} \\
    \hline
    \end{tabular}}
    \end{threeparttable}
    \end{center}
    \vspace{-25pt}
    \end{table}

\textit{Noisy Subject.} We implemented denoising approaches on noisy CEST data and 4-pool Lorentzian fitting successively. The denoising capability of IRIS allows for clear visualization of tumor shapes in CEST maps compared to T2w images, as shown in Fig.~\ref{nod5}. Moreover, in comparison to the normal tissue on the symmetrical counterpart, the denoised results using IRIS exhibit significantly higher APT intensity and lower rNOE effects in the tumor region, as the arrow indicated in Fig.~\ref{T2w_nod5}.
\vspace{-10pt}

\subsection{Ablation Studies}
\vspace{-3pt}
In this section, we discuss the selection of hyper-parameter $K$ and examine the differences in using various thresholding methods to solve Eq.~(\ref{eq:sub-prob}). All experiments in this section are implemented on synthetic phantom as it is completely noise-free. An optimal choice of $K$ value and thresholding method is crucial to balance preserving image details and reducing noise. Nevertheless, the mixture of noise and feature components after SVD makes simple truncation ineffective in achieving the desired denoising effect. As IRIS performs exceptionally in representing continuous functions in subspace by utilizing RegressionNet, it precisely captures feature-contributing components while disregarding those associated with noise. We replaced the thresholding method with the Median Filter and pre-trained DnCNN~\cite{zhang2017dncnn} model and compared them with IRIS as shown in Table~\ref{table:ablation study}. IRIS not only achieved the lowest MSE at $K=4$ but also maintained the most robust performance as more components were introduced. The vivo CEST mapping also yields the same conclusion (See \href{https://htmlpreview.github.io/?https://github.com/chuchen1206/IRIS_CEST-Denoising/blob/9e28e72a48ef32459931e468a6526df3808e03e0/IRIS_Appendix.html}{Appendix}).
    \begin{table}
    \begin{center}
    \caption{Assessment on the selection of $K$ and thresholding methods quantified in MSE~$(1\times10^{-4})$.}
    \vspace{-5pt}
    \label{table:ablation study}
    \begin{threeparttable}
    \setlength{\tabcolsep}{2mm}{
    \begin{tabular}{c c c c c c}
    \hline   \textbf{$K$} & 1 & 2 & 3 & 4 & 5 \\
    \hline  Truncation & 3.271 & 3.175 & 3.476 & 4.532 & 5.758 \\
            Median & 6.080 & 5.098 & 4.513 & 4.608 & 4.801 \\
            DnCNN~\cite{zhang2017dncnn} & 3.307 & 2.174 & 1.445 & 1.399 & 1.420 \\
            Current setting & \textbf{2.518} & \textbf{2.059} & \textbf{1.413} & \textbf{1.336} & \textbf{1.401} \\
    \hline
    \end{tabular}}
    \end{threeparttable}
    \end{center}
    \vspace{-25pt}
    \end{table}
\vspace{-12pt}
\section{Conclusion}
\vspace{-5pt}
In our work, we introduce the Implicit Regression in Subspace (IRIS) framework, a new unsupervised approach to denoise CEST MRI data. Through extensive evaluations from various angles, we have demonstrated that IRIS surpasses existing state-of-the-art methods. This framework refines a lightweight neural network via subspace regression, exhibiting exceptional resilience to varying degrees of noise. Notably, IRIS can produce high-sensitivity CEST maps even from data with minimal CEST effects, enhancing tissue contrast—particularly in critical areas like tumors. This advancement holds substantial promise for improving the diagnosis and treatment of tumors and furthering research into tissue characterization. Improving the denoising efficiency would be an interesting future work.

\vfill \pagebreak
\section{Acknowledgements}
This work is supported by HKRGC GRFgrants CityU1101120, CityU11309922, CRFgrant C1013-21GF, and HKRGC-NSFC Grant NCityU214/19.
\section{Compliance with Ethical Standards}
Approval of animal experiments was granted by the Animal Research Ethics Sub-Commitee of City University of Hong Kong (4. Sep. 2020/No. NSFC)

\bibliographystyle{IEEEbib}
\bibliography{refs}

\end{document}